# ASTROMETRIC DISCOVERY OF GJ 164B

Steven H. Pravdo
Jet Propulsion Laboratory, California Institute of Technology
306-431, 4800 Oak Grove Drive, Pasadena, CA 91109; spravdo@jpl.nasa.gov

Stuart B. Shaklan
Jet Propulsion Laboratory, California Institute of Technology
301-486, 4800 Oak Grove Drive, Pasadena, CA 91109, shaklan@huey.jpl.nasa.gov

Todd Henry
Department of Physics and Astronomy, Georgia State University
Atlanta, Georgia 30302-4106, thenry@chara.gsu.edu

and

G. Fritz Benedict
MacDonald Observatory, University of Texas at Austin
Austin, Texas 78712-1083, fritz@astro.as.utexas.edu

## ABSTRACT

We discovered a low-mass companion to the M-dwarf GJ 164 with the CCD-based imaging system of the Stellar Planet Survey (*STEPS*) astrometric program. The existence of GJ 164B was confirmed with *Hubble Space Telescope* NICMOS imaging observations. A high-dispersion spectral observation in *V* sets a lower limit of $\Delta m > 2.2$ mag between the two components of the system. Based upon our parallax value of $0.082 \pm 0.008$, we derive the following orbital parameters: $P = 2.04 \pm 0.03$ y, $a = 1.03 \pm 0.03$ AU, and $M_{total} = 0.265 \pm 0.020$ $M_\odot$. The component masses are $M_A = 0.170 \pm 0.015$ $M_\odot$ and $M_B = 0.095 \pm 0.015$ $M_\odot$. Based on its mass, colors, and spectral properties, GJ 164B has spectral type M6-8 V.

## 1. INTRODUCTION

M dwarfs are a large population of stars not yet systematically searched for low-mass companions, both planets and brown dwarfs. Of the ~100 extrasolar planets known (Marcy & Butler 2000, Butler et al. 2002, and references therein), only one system, GJ 876, has an M-dwarf primary (Marcy et al. 1998, Delfosse et al. 1998, Marcy et al. 2001). It is too early to conclude whether this low fraction is real or a selection effect.

We astrometrically search selected M-dwarfs for low-mass companions with the Stellar Planet Survey (*STEPS*, Pravdo & Shaklan 1996). As expected, we are detecting the largest mass companions with the largest signals first, and we expect to detect smaller mass companions down to a limit of ~$M_{Jupiter}$, as the baselines and sensitivities increase. Pravdo & Shaklan (2003) reported the detection of the low-mass companion to GJ 1245A, GJ 1245C (Harrington & Dahn 1984, McCarthy et al.1988, Harrington 1990), with a mass of ~0.07 $M_\odot$(Henry et al. 1999). Here we report the discovery of another low-mass companion around a nearby M star, GJ 164. GJ 164, also known at LHS 1642 and Ross 28, is 11-13 pc distant (van Altena, Lee, & Hoffleit 1995--YPC) with *V* = 13.5 (Weis 1996).

## 2. OBSERVATIONS, ANALYSIS, AND RESULTS

*2.1 STEPS Astrometry*

*2.1.1*
*STEPS Instrument and Program*

The *STEPS* instrument is mounted at the Cassegrain focus of the Palomar 200" telescope with an f/16 beam incident on the $LN_2$-cooled, 4096-squared, 15-μm-pixel CCD-camera. The dewar window is optically flat (< 1/30 wave peak-to-valley) to limit optical distortions and also serves as a filter in the 550-750 nm band to limit color effects. The plate-scale is 36 milliarcseconds (*mas*) per pixel resulting in a field 2.45´ on a side. We sum 2x2 physical pixels on chip to reduce the readout time to ~7 s via four output amplifiers after each 30-60-s exposure. This is acceptable because the ~1 arcsec point-spread functions (PSFs) are significantly oversampled.

*STEPS* began in 1997 with the current instrument and continues with 30 program stars. We observe ~12 nights/year spread over most of the year except March-June, when our targets are inaccessible. Seasonal patterns of bad weather have greatly affected the temporal sampling. The *STEPS* targets are all nearby M dwarfs, fainter than *V*~12 to avoid saturation, and at low galactic latitude for a sufficient number (≥6) of reference stars in the field. Measurement noise is comprised of Poisson noise and systematic noise due to color differences between the target and reference stars (differential chromatic refraction [DCR]) and variability in the atmosphere, telescope optics, camera electronics, and CCD geometry. These are discussed in the Appendix (see also Pravdo & Shaklan 1996). The current noise floor is ~1 *mas*, somewhat above the theoretically expected value of ~0.5 *mas*.

*2.1.2*
*Astrometric Modeling*

We use in-frame relative astrometry to find stellar companions via the wobble of the primary around the center of mass. The model is a two-dimensional linear model of the form:

$$x' = c_1+c_2*x+c_3*y \qquad (1)$$
$$y' = c_4+c_5*x+c_6*y$$

where (x , y) are raw coordinates, (x′, y′) are fitted coordinates, and the $c_i$ are the six coefficients of the fit.

The signal of a companion is first revealed as a periodic residual after the much larger motions due to parallax and proper motion are removed from each target motion.

We then use the non-linear Marquardt routine (Press et al 1989) to fit the total motion to a model with 14 possible parameters: seven orbital elements, two proper motions, one parallax, two celestial positions at an epoch, one radial velocity, and a parameter that depends upon the light ratio in the "high-mass" branch ($M_{comp} \geq 0.08\ M_\odot$), or upon the mass ratio in the "low-mass" branch ($M_{comp} < 0.08\ M_\odot$). In practice the mean anomaly is set to zero at the derived epoch, and the celestial positions, and radial velocity are fixed since the model is insensitive to small changes in their values, leaving 10 free parameters. The multi-dimensional χ-squared space is quite complex and we use a Monte Carlo program that explores the parameter space in conjunction with the Marquardt routine. One sigma uncertainties are calculated with the method of Lampton, Margon, & Bowyer (1976) for parameter estimation in multi-dimensional models.

We observe the movement of the center of light around the center of mass: the photocentric orbit. In the limit where the companion light is negligible this is the same as the Keplerian orbit of the primary. However, when the companion makes a significant contribution to the light via self-luminosity, the orbit of the photocenter is reduced, and approaches the limit of zero motion when the companion and the primary have the same mass and light. Therefore, for a given observed motion, there are two possible solutions. If we detect a small motion of the primary it could be due to either a relatively small-mass, non-luminous secondary or to a relatively large-mass, luminous secondary. Distinguishing between the two possibilities requires imaging or spectroscopic information.

The ratio of the photocentric and Keplerian orbits, $\alpha/a$, is (e.g., McCarthy et al. 1988):

$$\alpha/a = f - \beta \qquad (2)$$

where $f$ is the fractional mass of the secondary, $\beta$ is the fractional light, $a$ is the semi-major axis and $\alpha$ is the photocentric radius. We use the visible mass-luminosity relation (MLR, Henry et al. 1999) to relate the absolute luminosity of each component to its mass. As the mass and absolute luminosity of the secondary decreases, $\beta$ becomes negligible compared with $f$ in equation (2), and the MLR also becomes inapplicable. This limit is reached at $0.08\ M_\odot$. Figure 1 plots $f$, $\beta$, and $\alpha/a$ as a function of the secondary mass for this system, and also demonstrates that for each $\alpha/a$ value, aside from maximum, ~0.31, there are two possible secondary masses.

*2.1.3 GJ 164 Parallax and Proper Motion*

We determine parallax and proper motions relative to our reference frame. Our values may differ from the absolute values because the reference frame is not infinitely distant. The GJ 164 frame contains 9 reference stars with an average in-band intensity of ~1% of GJ 164 implying an average distance 10 times farther. Additionally, in all cases, the reference stars are bluer than GJ 164 implying earlier spectral types, higher luminosities, and even larger distances. The effects of absorption and reddening at this galactic latitude, ~1°, must also be taken into account. We use herein a standard correction as described in the following section. However, we use our relative values in our astrometric fits because they are appropriate for removing these motions in our data.

In determining the final system parameters we allow for an absolute parallax value that differs from our fitted value.

The literature contains several values for the both the proper motion and parallax of this star. The Revised LHS catalog (Bakos, Sahu, & Nemeth 2002) has –284 *mas* y$^{-1}$ in RA and –854 *mas* y$^{-1}$ in Decl., the LHS catalog (Luyten 1979) has -366 *mas* y$^{-1}$ in RA and –833 *mas* y$^{-1}$ in Decl., and, Heintz (1993) gives –320 *mas* y$^{-1}$ in RA and –811 *ma*s y$^{-1}$ in Decl.

Three previously determined trigonometric parallax values are: 0.063 ± 0.0104 (van Maanen 1931), 0.0782 ± 0.0082 (Hershey 1980), and 0.0868±0.0067 (Heintz 1993). These are combined to yield a weighted average for the absolute parallax of 0.0839±0.0087 (YPC). We suggest that the variation among the prior results was caused in part by an interaction with the proper motion values that are also significantly divergent, with the underlying cause, the time-dependent, and unaccounted-for contribution of the companion.

*2.14 GJ 164 results*

We have now observed GJ 164 with *STEPS* from Dec. 19, 1997 to Feb. 17, 2004, or more than 6 years. During that time GJ 164 moved ~2000 *mas* in RA and ~5000 *mas* in Decl. due to its proper motion. If we fit our data without a companion we get a parallax of 0.073 and proper motions of –322 *mas* y$^{-1}$ in RA and –808 *mas* y$^{-1}$ in Decl., the latter consistent with prior results. The addition of GJ 164B to the model fit changes our results only slightly.

The addition of a companion to the total motion model significantly reduces the residuals in the fit from ~18 *mas* to ~2 *mas*. The ~2 *mas* residual is larger than the ≤1 *mas* residuals we achieve in other data sets and is due to either other real motion in the system or systematic error. Tables 1 and 2 give our best-fit values for the orbital parameters, relative parallax, and proper motion. Figure 2 shows a fit of the residuals to a model with a companion in a 2.04-y period after parallax and proper motion subtraction. A ~4-y orbital period also fits these data because of the uncertainty introduced by the ~1000 day data gap but is less likely to be correct based upon other information (see below). For GJ 164 the low- and high-mass branches of the model merge since the orbit ratio ($\alpha/a$) is near ~0.31 (Figure 1). In our fits, $\alpha/a = 0.289 \pm 0.017$, $f = 0.32 \pm 0.06$, and $\beta = 0.046 \pm 0.044$.

Our relative parallax, 0.071 (+0.004,-0.002), falls among the prior results. To correct this to the absolute parallax, we add ~2 *mas*, typical for stars of this magnitude and galactic latitude (YPC, Figure 2), yielding 0.073 (+0.004,-0.002). However, our imaging and spectroscopic results, if we assume that GJ 164A is well described by the visible and near-infrared MLRs, imply a larger value for the parallax (see §3).

*2.2 Hubble Space Telescope Imaging*

We performed observations with the NICMOS 1 camera (NIC1) of *Hubble Space Telescope (HST)* on Dec. 23, 2003 and Feb. 14, 2004, using the F108N and F190N filters at a fixed, arbitrary roll angle on the first date, and the F108N, F164N, and F190N filters at a fixed roll angle, 45° from the first roll angle, on the second date, for a total of five observations. The observations were dithered in a spiral pattern with 0.445'' step size to smooth pixel effects. We searched for a faint companion to GJ 164A in two ways. First we subtracted the pipeline-processed mosaicked images with the same filter and different roll angles to eliminate persistent effects such as the image of the primary and any other azimuthally constant features. Second we used the PSFs in TinyTim (http://www.stsci.edu/software/tinytim/tinytim.html) to create a model of two point sources representing components A and B. We used this model to fit the five observations with relative flux, separation, and position angle (PA) as the free parameters. Figure 3 shows the resulting images after subtraction of component A and Table 3 lists the results. We include in Table 1 an approximate guide to the relative fluxes of the components in *JHK*, acknowledging that the narrow HST filter bands are only rough approximations; with F164N perhaps the best match as it is well centered in *H*.

The most important result seen in all 5 images is the confirmation of the existence of the astrometric companion, GJ 164B. The companion is to the southeast in the December images and to the east in February. GJ 164B cannot be a stationary background object because it would have appeared to move northeast as GJ 164A's proper motion carried it southwest. If it were stationary it would be 84 *mas* north of GJ 164A in the second observation, rather than the measured 15 *mas* south.

The *HST* positional data were represented in the *STEPS* format, i.e. an RA and Decl. positional offset at a Julian Date, and we fit the astrometric model to the *STEPS* and *HST* data simultaneously. The 2.04-y astrometric orbit is a significantly better fit than a 4-y model (2 *mas* compared with 3.4 *mas* residuals), resulting in values of 75 mas at 114º PA, and 86 mas at 101º PA, in good agreement with the *HST* observations (Table 3).

Table 3 also shows the photometry for the three bands. The light ratios are about 6-7 to 1 in each band with uncertainties of ~10%.

*2.3 Spectroscopy*

We searched for evidence that GJ 164 is a double-lined spectroscopic binary (SB2). The observations were made on February 12, 2004 with the Sandiford Cass Echelle spectrograph on the McDonald Observatory 2.1-m telescope. We used the IRAF task *fxcor* to cross-correlate the spectrum with a template high-signal-to-noise spectrum of the M2.5V J dwarf Gl 623 (Henry et al. 1994), obtained at the 9.2-m Hobby-Eberley Telescope. We consider Gl 623 to be an effectively "non-binary" template because its two components differ by ~5.3 magnitudes (Henry et al. 1999). Figure 5 shows a sample correlation function for one of the 22 echelle orders. It appears very sharp and symmetrical. A weighted average of the velocities from each of 22 orders yields a radial velocity, Vr = -29.9 ± 0.3 km s$^{-1}$ (heliocentric), where the velocity uncertainty in the ephemeris of our binary template has been included. We then attempted to de-blend the eight cleanest correlation peaks, but found no evidence of duplicity. We estimate that the *V* magnitude difference between components is at least 2.2. We base this limit on previous successful analysis of Wolf 1062 (Gl 748), an SB2 with Δm = 1.8 (Benedict *et. al* 2001), and in an ongoing investigation marginal detection of the two components of Wolf 922 (Gl 831, Δm =2.1, Henry et al. 1999) only at perihelion (largest ΔV$_r$). This would restrict the mass of the companion to be less than 0.11 M$_\odot$, confirming the astrometric upper limit (Table 2).

Figure 4 also shows the location in the orbit when the spectroscopic observation was made (asterisk on plot). At that time the estimated ΔV$_r$ ~13 km s$^{-1}$. Current extra-solar planet-finding radial velocity experiments should detect the motion of the *V*~13 primary in this orbit.

### 3. DISCUSSION

The GJ 164 orbit as determined from astrometry is shown in Figure 4 where we have taken the measured photocentric orbit of the primary and transformed it into an orbit that shows the motion of the secondary. We superimpose on the plot the points in the orbit when we had HST and spectral observations. There is consistency between the astrometry and the imaging with respect to the position angles, the direction of motion, and the separation between the components.

Our imaging and spectroscopic measurements and the MLRs in *H* (Henry & McCarthy 1993) and *V* (Henry et al. 1999) bands allow us to infer both the primary mass and the parallax, assuming that the primary is neither over- nor under-luminous in *H* compared with *V*. Since the *HST* F164N measurement is narrow and centered in the *H* band, we take its ratio to be approximately equal to the *H* ratio. The F108N band is contained in *J* and the F190N band is contained in *K* band but both are off-center so the ratios are probably less representative. Nevertheless we use them for zeroth order estimates of the *JHK* Δm (Table 1). We obtain upper limits to the primary mass (as a function of parallax) by assuming that all the visible light comes from the primary. Likewise a lower limit is obtained by using the upper limit to the *V* Δm (§2.3). Next we calculate *H* magnitudes for the primary from the *H* MLR, for these mass limits. The difference between the calculated *H* luminosity of the primary and the 2MASS value tells us the *H* magnitude of the secondary and thus, the secondary to primary ratio. This ratio is assumed equal to the F164N ratio. Parallax values <0.086 are not allowed as they would result in *H* ratios that are too high. Similarly, parallax values > 0.090 are not allowed, as they would result in *H* ratios that fall below the F164N ratio. Figure 6 shows the *V*-based limits (solid lines) and the *H*-based limits (dotted lines). The overlapping region constrains the parallax to be 0.086 ± 0.002 and the GJ 164A mass to be 0.163 ± 0.004 $M_\odot$, without considering the uncertainty in the MLR itself (≥20%, see following). The F164N ratio and the *H*-based MLR then constrain $M_B \leq 0.092$ $M_\odot$ with an uncertain lower limit below 0.08 $M_\odot$ since the MLR is not valid in that range.

If the parallax is 0.086-0.090 rather than our corrected value of 0.073, then the masses of the objects and the size of the orbit are smaller, while the period remains the same. The net effect is to reduce the orbital effect of the secondary by either lowering its mass or increasing its light. For parallaxes above 0.082, the central portion of the secondary mass range, $M_B \sim 0.075\text{-}0.10$ $M_\odot$, is a slightly worse fit than before (3 rather than 2 *mas* residuals). However, we do not believe that $M_B < 0.075$ $M_\odot$, based upon the inferred *JHK* values. The other possibility, $M_B > 0.10$ $M_\odot$, is inconsistent with $M_B < 0.092$ $M_\odot$ based upon the H magnitudes analysis above. This inconsistency would be eliminated, however, if the *H* ratio between the components were 0.25 (dashed line in Fig. 6) rather than 0.17 ± 0.01 or if the primary mass were 0.177 $M_\odot$ rather than 0.167 $M_\odot$ corresponding to a difference in the *V* MLR of 0.28 mag in $M_V$. This last possibility is not unreasonable since 0.28 < 0.5-1.0-mag spread expected for stars of this mass with varying metallicity (Baraffe et al. 1998). Our estimate for the absolute parallax is thus $0.082 \pm 0.008$. The primary and secondary mass estimates are then $M_A = 0.170 \pm 0.015$ $M_\odot$ and $M_B = 0.095 \pm 0.015$ $M_\odot$.

The likely stellar type of GJ 164B is M6-M8, with the uncertainty due mostly to the parallax. A similar star might be GJ 644C (Henry et al. 2004), classified as M7.0 V with $M_J$, $M_H$, $M_{Ks}$ = 10.73, 10.15, 9.77, respectively, while GJ 164B has 10.62, 10.42, 9.63 (Table 1).

There has been no sensitive X-ray observation of GJ 164; an upper limit of ~8 x $10^{27}$ erg s$^{-1}$ can be inferred during the epoch of the *ROSAT* survey (Hünsch et al 1999). This does not place significant constraints on whether GJ 164 is "X-ray active" although the lack of significant Hα emission is associated with low coronal X-ray emission (Doyle 1989) and with older, slower rotators. An age estimate based on the *V-I$_C$* color method (Hawley et al 1999, Gizis, Reid, & Hawley 2002) yields ~2.5 billion years.

### 4. CONCLUSION

We have astrometrically discovered a low-mass companion to GJ 164A and confirmed the discovery with imaging. Future observations will reduce the uncertainties in the GJ 164B mass, type, and age, and will allow us to further explore the properties of stars in this difficult to observe mass region.

### APPENDIX

The data reduction process begins by extracting square regions containing the target and reference stars from the raw frames and organizing them into a single file. If desired, the corresponding regions from flat-field files are also extracted for use in the centroiding step.

Next, the positions for all stars are determined. Our algorithm measures the cross-correlation of the reference stars relative to the target. The cross-correlation is determined by the weighted slope of the phase of the Fourier Transform (FT) of the images after summing them into horizontal and vertical distributions. The algorithm has several advantages compared to traditional techniques. Standard centroids and Gaussian fits are sensitive to the background level and shape of the PSF. Our algorithm is insensitive to the background level, robust against changes in the shape of the PSF, and maintains SNR comparable to matched filtering.

We form a preliminary astrometric solution after centroiding by fitting a conformal 6-term (3 per axis) transformation for each CCD frame to a reference frame. The transformation is then applied to the target star, allowing the target star position to be measured relative to the surrounding reference stars.

An automated frame-editing program searches for frames whose astrometric noise is above a user-defined threshold. Generally the threshold is very high (20 sigma) because the major cause of unusable data is missing reference stars or the selection of the wrong star. After frame removal, the conformal transformation is re-run to form an intermediate astrometric solution.

Because the stars do not all have the same effective color, they are affected by differential chromatic refraction (DCR). The DCR amplitude is as large as ~10 *mas* for our bandpass and range of zenith distances. The effect is proportional to the tangent of the zenith angle, and manifests itself as a linear drift in RA and a parabolic shift (relative to the meridian position) in Decl. The relative DCR coefficient for each star is determined by fitting the RA drift.  A single coefficient is determined for each star as the weighted average of the nightly coefficients.

The centroid positions are then adjusted to account for DCR and the conformal transformation is re-run. All observations are made within 1.5 hours of meridian transit to minimize chromatic effects. At this point, the motion of the target star relative to the reference frame is known and plotted.  Nightly statistics are computed and the mean position and time of observation are written to a file.

The final processing step is to fit the motion of the target star to a model that contains the parallax, proper motion, and radial velocity of the target stars and the effect of any companion.  Note that we determine and use the relative, rather than absolute, proper motions and parallaxes. In practice the model is insensitive to radial velocity, but we input its fixed value when known. We use the USNO subroutine ASSTAR to compute the astrometric place of the star from its mean place, proper motion, parallax, and radial velocity, and the NAIF subroutine CONICS to determine the state (position, velocity) of an orbiting body from a set of elliptic orbital elements.

## ACKNOWLEDGMENTS

The research described in this paper was performed in part by the Jet Propulsion Laboratory, California Institute of Technology, under contract with the National Aeronautics and Space Administration. We performed observations at Caltech's Palomar Observatory and acknowledge the assistance of the staff. This research has made use of the NASA/IPAC Infrared Science Archive, which is operated by the Jet Propulsion Laboratory, California Institute of Technology, under contract with the National Aeronautics and Space Administration. This research has made use of the SIMBAD database, operated at CDS, Strasbourg, France, and of NASA's Astrophysics Data System Abstract Service. This publication makes use of data products from the Two Micron All Sky Survey, which is a joint project of the University of Massachusetts and the Infrared Processing and Analysis Center/California Institute of Technology, funded by the National Aeronautics and Space Administration and the National Science Foundation. We also acknowledge use of the NStars Database. We thank the referee W. van Altena for useful suggestions.

## Table 1. GJ 164 Properties

|  | Literature | This work | |
|---|---|---|---|
| RA (FK5 2000/2000)[a] | 04$^h$ 12$^m$ 58.94$^s$ | - | |
| Decl. (FK5 2000/2000)[a] | 52º 36´ 40.7´´ | - | |
| Proper Motion (arcsec) | 0.910$b$ | 0.871 ± 0.002 | |
| PA (deg.) | 203.7[b] | 201.9 ± 0.1 | |
| Relative Parallax | - | 0.071 (+0.004,-0.002) | |
| Absolute Parallax | 0.0839 ± 0.0087[c] | 0.0820 ± 0.008 | |
| (Π=0.082 for following) |  | Primary | Secondary |
| M$_V$ | 13.10[d] | <13.23 | >15.43 |
| M$_J$ | 8.37[e] | 8.48 | 10.62 |
| M$_H$ | 7.84[e] | 7.92 | 10.42 |
| M$_{Ks}$ | 7.51[e] | 7.65 | 9.63 |
| Spectral Type | M4.5 V[f] | M4.5 V | M6-8 V |

[a]Bakos, Sahu, & Nemeth (2002)
[b]Luyten 1979
[c]van Altena, Lee, & Hoffleit 1995
[d]Weis 1996
[e]2MASS
[f]Reid, Hawley, & Gizis 1995

## Table 2. Orbital Parameters

| Parameter | From Astrometry | Plus Imag. & Spect. |
|---|---|---|
| Period (y) | 2.04 ± 0.03 | - |
| Semi-major axis (AU) | 1.02 ± 0.05 | 1.03 ± 0.03 |
| Eccentricity | 0.35 ± 0.22 | - |
| Inclination | 57 (+13,-17) | - |
| Longitude of ascending node (deg) | 12 ± 15 | - |
| Argument of periapse (deg) | 133 (+45,-20) | - |
| Epoch | 1998.9 ± 0.2 | - |
| Total mass | 0.255 ± 0.030 | 0.265 ± 0.020 |
| GJ 164A mass | 0.166 ± 0.020 | 0.170 ± 0.015 |
| GJ 164B mass | 0.089 ± 0.020 | 0.095 ± 0.015 |

## Table 3. HST Results

| Quantity | Value | | |
|---|---|---|---|
| Filter | **F108N** | **F164N** | **F190N** |
| Flux ratio | 0.14±0.02 | 0.17±0.01 | 0.17±0.01 |
| Date | **Decl. 12, 2003** | **Feb 14, 2004** | |
| Separation (*mas*) | 75 ±1 | 88 ±3 | |
| PA (deg) | 116±3 | 100±3 | |

FIGURE CAPTIONS

1. We give the fractional mass ratio, $f$ (solid line), fractional light ratio, $\beta$ (dashed line), and photocentric-to-relative orbit ratio, $\alpha/a$ (dotted line), as a function of the GJ 164B mass. The range of acceptable astrometric fits is shown by "astrometry limits."

2. RA and Decl. relative motion of the GJ 164 photocenter measured with *STEPS*. The points show the data (RA-filled diamonds, Decl.-open squares) and the lines (RA-solid, Decl.-dashed) show the 2.04-y model with 0.10 $M_\odot$ secondary. Errors on the points are 2 *mas*.

3. Five HST images with the TinyTim point-spread function (PSF) of GJ 164A subtracted. The top two are from the first date (initial roll angle and two filters), and the bottom three are the second date, 53 days later (second roll angle and three filters). The contours represent light levels including GJ 164A.

4. The 2-y astrometric orbit of GJ 164B around GJ 164A (filled square). The filled diamonds are the positions of the companion in 30-day intervals. We also show the positions at the times of the HST (open squares) and spectroscopic (asterisk) observations.

5. Cross-correlation function obtained for GJ 164, using a Gl 623 template. The spectral range for this particular order was 547 - 553 nm. The correlation peak is indistinguishable from that of a single star.

6. GJ 164A mass and parallax limits are shown. The solid lines are the maximum mass based upon the *V* MLR, and the minimum mass based upon the observed *V* Δm. The dotted lines are the limits imposed on the mass by the *HST* F164N light ratio and the *H* MLR. The dashed line corresponds to a GJ 164B/GJ 164A *H*-luminosity ratio of 0.25.

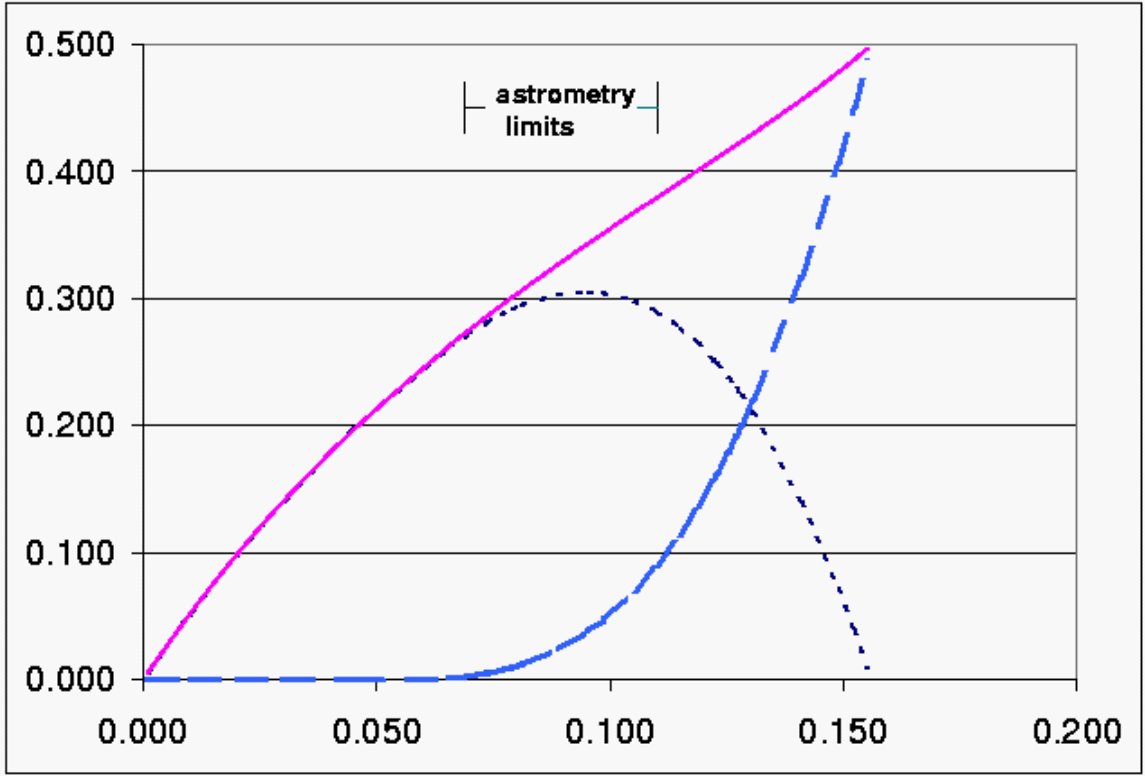

**Figure 1**

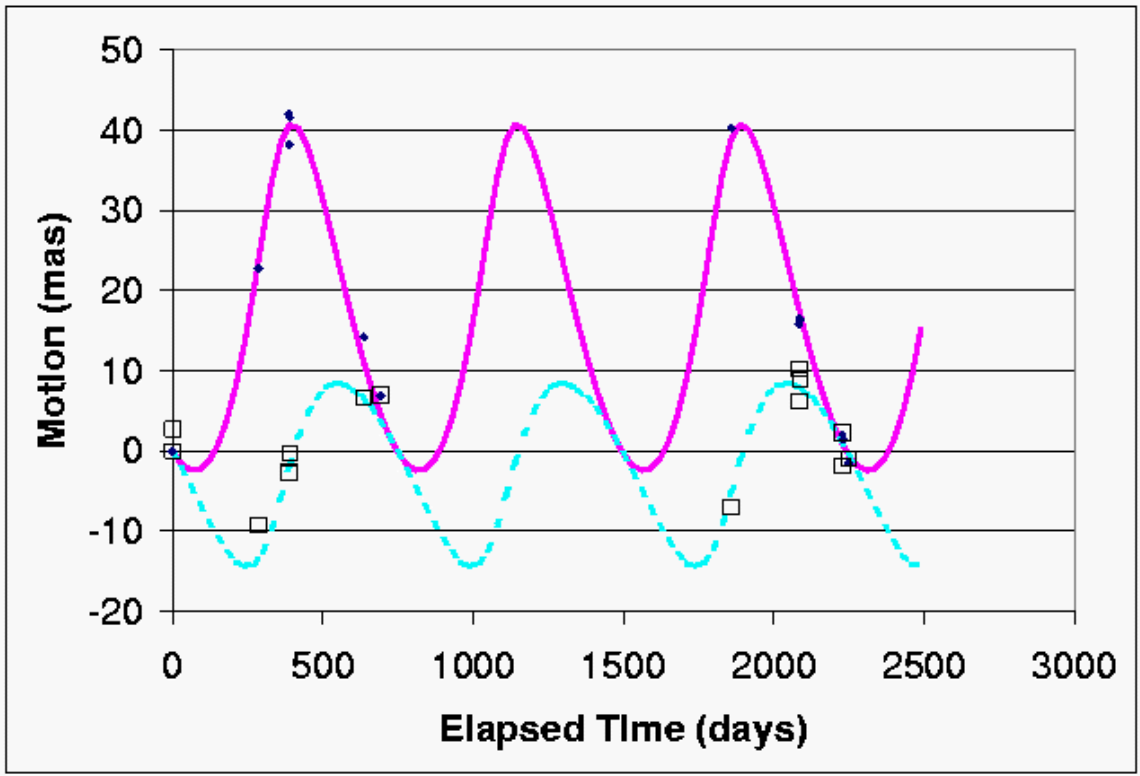

**Figure 2**

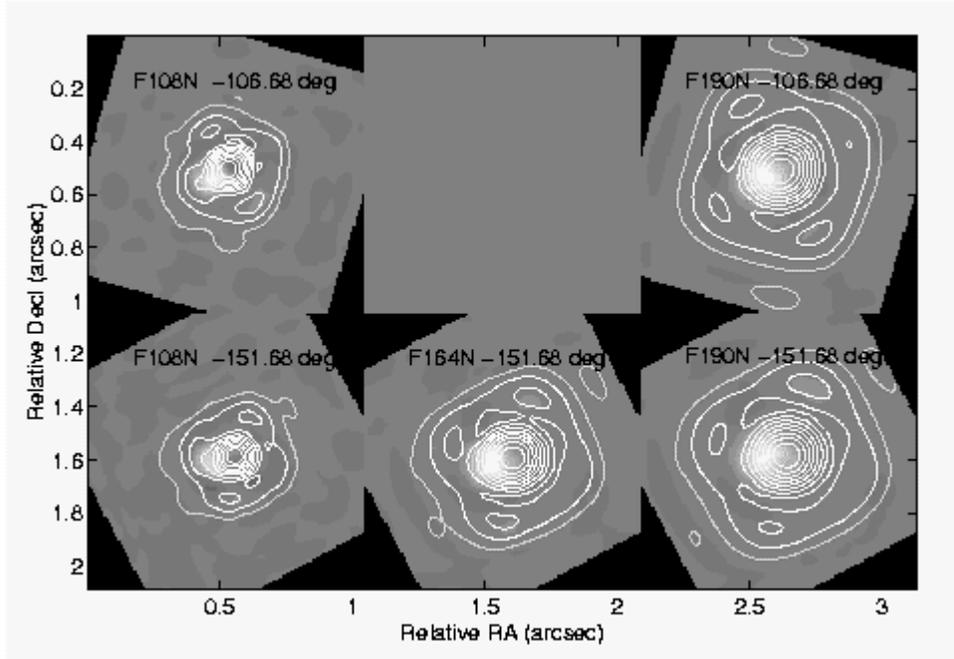

**Figure 3**

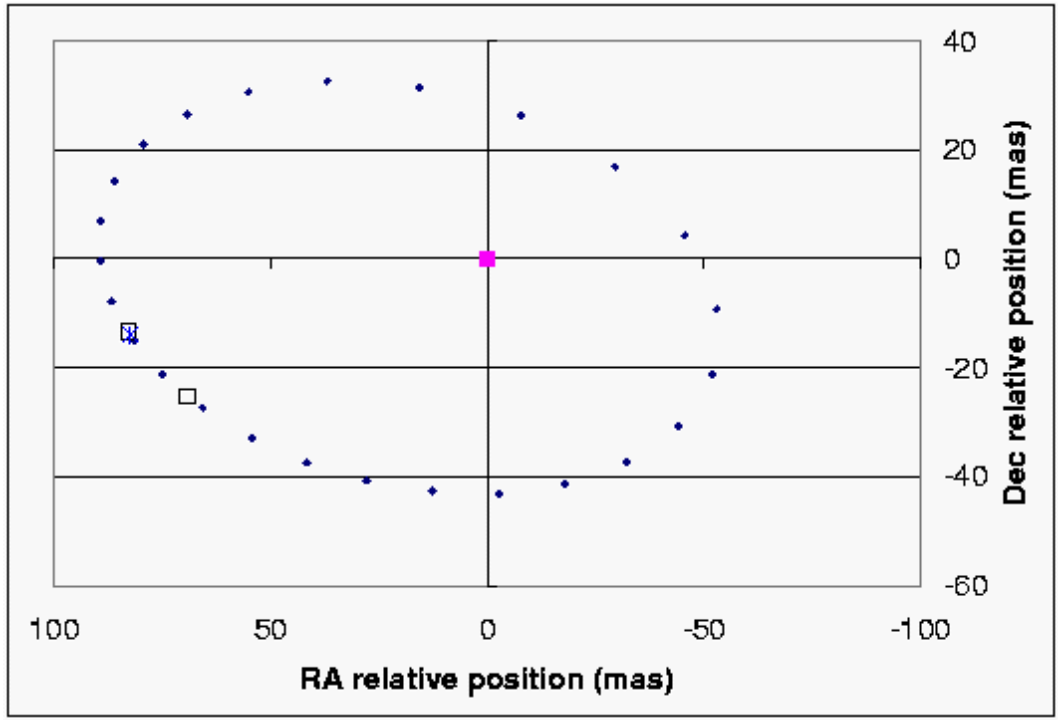

**Figure 4**

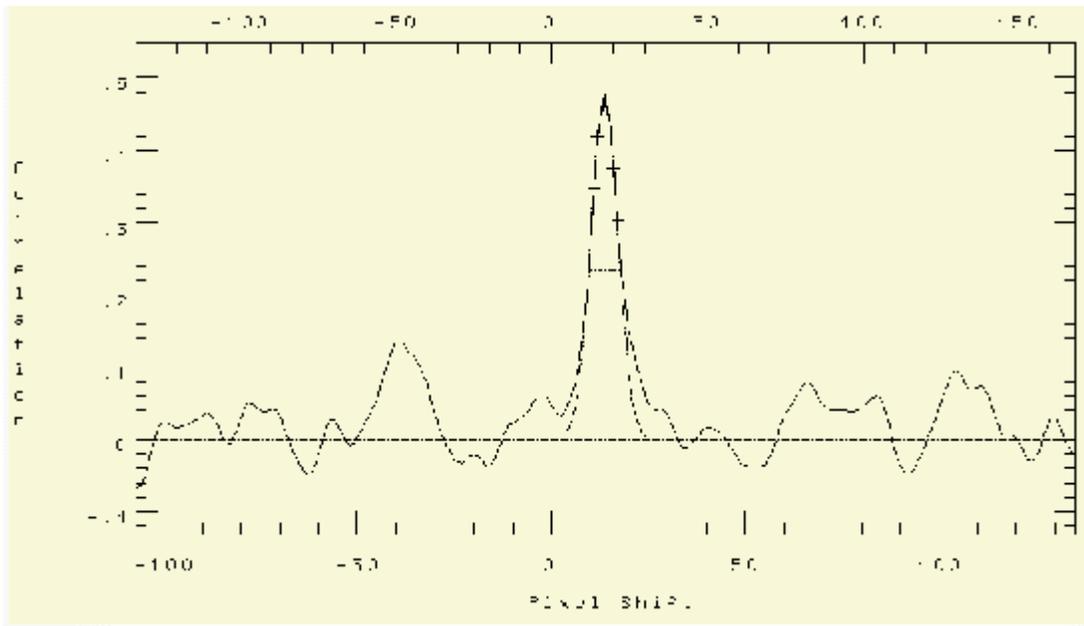

**Figure 5**

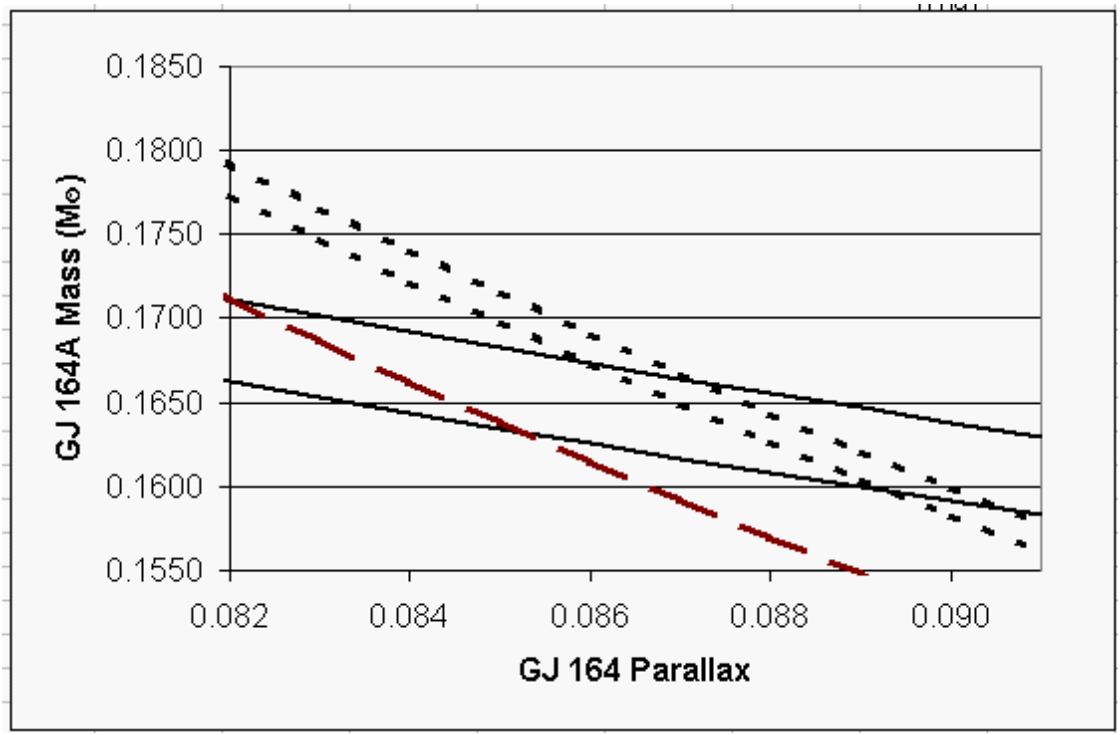

**Figure 6**